\documentclass[usegraphicx,usenatbib,useAMS]{mn2e}

\usepackage{times,amssymb}

\newcommand\ion[2]{\ensuremath{\mbox{#1\,{\sc#2}}}}
\newcommand{\sneia}{SNe~Ia}
\newcommand{\snic}{SN~Ic}
\newcommand{\sneic}{SNe~Ic}
\newcommand{\lam}{\ensuremath{\lambda}}
\newcommand{\nifs}{\ensuremath{^{56}\rm{Ni}}}
\newcommand{\msun}{\ensuremath{M_{\odot}}}
\newcommand{\kms}{\ensuremath{\rm{km\,s}^{-1}}}
\newcommand{\ergs}{\ensuremath{\rm{erg\,s}^{-1}}}
\newcommand{\bv}{\ensuremath{B\!-\!V}}
\newcommand{\ebv}{\ensuremath{E(\bv)}}
\newcommand{\ubvri}{\ensuremath{U\!BV\!R\!I}}
\newcommand{\jhk}{\ensuremath{J\!H\!K}}
\begin{document}

\title[The properties of SN~1994I from spectral models]{%
The properties of the ``standard'' type~Ic supernova 1994I from spectral
models}

\author[D.~N.~Sauer et al.]{%
D.~N.~Sauer,$^{1,9}$\thanks{email: {\tt sauer@ts.astro.it}}
P.~A.~Mazzali,$^{2,1,4,7,9}$ J.~Deng,$^{3,4,9}$ S.~Valenti,$^{5,6}$ K.~Nomoto,$^{4,7,9}$ \newauthor
and A.~V.~Filippenko$^{8}$\\
$^{1}$INAF, Osservatorio Astronomico di Trieste, Via G.B. Tiepolo, 11, I-34131 Trieste, Italy\\
$^{2}$Max-Planck-Institut f\"ur Astrophysik, Karl-Schwarzschild-Str.  1, 85741 Garching, Germany \\
$^{3}$National Astronomical Observatories, Chinese Academy of Sciences, 20A Datun Road, Chaoyang District, Beijing 100012, China\\
$^{4}$Department of Astronomy, School of Science, University of Tokyo, 7-3-1 Hongo, Bunkyo-ku, Tokyo 113-0033, Japan\\
$^{5}$European Southern Observatory, Karl-Schwarzschild-Str. 2, 85748 Garching, Germany \\
$^{6}$Physics Department, University of Ferrara, I-44100 Ferrara, Italy\\
$^{7}$Research Center for the Early Universe, School of Science, University of Tokyo, Tokyo, Japan\\
$^{8}$Department of Astronomy, University of California, Berkeley, CA 94720-3411, USA\\
$^{9}$Kavli Institute for Theoretical Physics, University of California, Santa Barbara, CA 93106-4030, USA
}

\maketitle

\begin{abstract}
   The properties of the type~Ic supernova SN~1994I are re-investigated.
   This object is often referred to as a ``standard SN~Ic''  although it
   exhibited an extremely fast light curve and unusually blue early-time
   spectra. In addition, the observations were affected by significant
   dust extinction. A series of spectral models are computed based on the
   explosion model CO21 (Iwamoto et~al. 1994) using a Monte Carlo
   transport spectral synthesis code.  Overall the density structure and
   abundances of the explosion model are able to reproduce the
   photospheric spectra well.  Reddening is estimated to be $\ebv=0.30\,$mag,
   a lower value than previously proposed.  A model of the nebular
   spectrum of SN~1994I points toward a slightly larger ejecta mass than
   that of CO21. The photospheric spectra show a large abundance of
   iron-group elements at early epochs, indicating that mixing within the
   ejecta must have been significant. We present an improved light curve
   model which also requires the presence of {\nifs} in the
   outer layers of the ejecta.
\end{abstract}

\begin{keywords}
radiative transfer -- line: formation --- line: identification --- supernovae: general --- supernovae: individual (SN~1994I) --- gamma rays: bursts
\end{keywords}

\section{Introduction}

With the proposed connection between core-collapse supernovae of type~Ic
({\sneic}) and long-duration gamma-ray bursts (GRBs; see \citealt{piran99} for
a review), interest in {\sneic} has recently increased. The proposed picture of
{\sneic} is the collapse of the core of a massive star which shed its hydrogen
and helium envelope prior to the explosion. In contrast to type~Ia supernovae,
core-collapse supernovae exhibit a very diverse appearance. The sub-class of
type~Ic supernovae which have been occasionally connected to GRBs are
characterised by large kinetic energies in their ejecta and are often referred
to as ``hypernovae'' or broad-lined supernovae. The progenitors of this class
of objects are proposed to be very massive stars with masses above $25\msun$
and are likely to form black holes after their collapse, while {\sneic}
originating from lower mass stars probably leave a neutron star behind
\citep[e.g.,][]{nomoto02}.

In order to understand the properties of {\sneic} in general and how ``normal''
{\sneic} relate to the class of ``hypernovae,'' a systematic analysis of
well-observed objects is needed. SN~1994I is regarded as a prototypical
{\snic}. It has extensive observational coverage
\citep{yokoo94,filippenko95,richmond96,clocchiatti96}, and it has been the
subject of various theoretical studies. Both synthetic light curves and spectra
have been published previously
\citep{nomoto94a,iwamoto94,baron96,millard99,baron99}; however, so far no
consistent picture of the properties of this object has emerged.

SN~1994I is also important because it seems to represent the branching point of
the model sequence of core-collapse supernovae proposed by \citet{nomoto03a}.
Therefore, it could provide the key for a more systematic understanding of the
evolution of the progenitors of the various flavours of core-collapse supernovae.

\citet{iwamoto94} used a low-mass model (CO21) to fit the light curve of
SN~1994I. Their model has only $\sim\!1\msun$ of CO-rich ejecta, suggesting
that SN~1994I was the stripped C-O core with a mass of $\sim\!2.1\msun$
originating from a $\sim\!4\msun$ He-core. This corresponds to a star with a
main sequence mass of $\sim\!15\msun$.  To achieve such a thorough stripping,
repeated interaction with a binary companion must be envisaged.

In this work we present a series of spectral models for the early-time spectra
of SN~1994I based on the explosion model CO21 by \citet{iwamoto94}; see also
\citet{nomoto94a}. This explosion model provided an excellent fit to the light
curves of SN~1994I. However, light curve models alone are not able to constrain
the explosion parameters uniquely because the light curve shape depends on a
combination of kinetic energy and ejecta mass. The peak luminosity is
constrained by the amount of {\nifs} synthesised during the explosion. To
determine the total luminosity,  it is necessary to have reliable values for
parameters such as reddening and distance. In addition, the bolometric
correction used to infer the luminosity has to be derived from the shape of the
spectral energy distribution.   Using the highly parametrized spectral code
SYNOW \citep[][and references therein]{fisher99}, \citet{millard99} provided
line identifications for some of the spectra of SN~1994I.  \citet{baron99} have
calculated non-local-thermodynamic equilibrium (non-LTE) models for a few
spectra based on the outcome of CO21 and concluded that this model is not able
to reproduce the observed spectra in detail. They suggested that a higher mass
model would be required to match the line features better.

\section{Spectral models}
\label{sec:ph-models}

We used the observed spectra published by \citet{filippenko95} which have been
flux calibrated against the photometry of \citet{richmond96}.  Our series of
photospheric models covers the observations from April 4 to May 4, 1994. (UT
dates are used throughout this paper.) For all models the explosion date was
set to be March 27, 1994 (JD 2\,449\,439), corresponding to a rise time to $B$-band
maximum of $12\,$d \citep{iwamoto94}. In what follows the epochs are given in
days relative to this date. The $B$-band maximum of SN~1994I occurred on April
7 (JD 2\,449\,450) \citep{richmond96}. We have excluded the first spectrum
(April 2) in \citet{filippenko95} because the wavelength coverage of this
observation is too narrow for a conclusive model. Spectra later than May 5
already show too much nebular emission to be treated with a photospheric
approach. In addition, we used a spectrum from April 3 that was taken at the
Multiple Mirror Telescope (MMT) by Schmidt et al. (1994, priv. comm.).

The early-time spectral models have been computed employing the method which
was first proposed by \citet{abbott85} and further developed by
\citet{mazzali93b}, \citet{lucy99}, and \citet{mazzali00}.  The method employs
a Monte Carlo simulation of the line transfer based on the Sobolev
approximation and includes a line-branching scheme.  At the end of the Monte
Carlo calculation the emergent spectrum is derived from a formal integral
solution on the basis of the source functions that can be extracted from the
Monte Carlo simulation \citep[see][]{lucy99}.

For the model computation the ejected material is divided into an outer region,
where the true continuum opacity is assumed to be negligible, and an inner
region which is not treated explicitly. At the dividing boundary of the
computational area a thermal blackbody continuum is imposed with a temperature
that is adjusted according to the luminosity of the emergent radiation at the
outer boundary. Note that this temperature has to be determined in an iterative
process because it is {\it a priory} not clear what fraction of the radiation
is scattered back into the core region. This constitutes the effect of line
blanketing where the back-scattered radiation causes the matter to heat up. The
ejecta are treated in spherical symmetry. At this stage we assume a homogenised
composition for the entire computational volume.

To fit the early-time spectra we used the density structure from the CO21
explosion model \citep{nomoto94a,iwamoto94}. The model is expanded homologously
to match the desired epoch. The composition was adjusted in order to obtain a
good fit to the observed spectrum.  The abundance distribution from the
nucleosynthetic yields of the CO21 model averaged over velocity space above
each epoch's photosphere served as a guideline. We adopt a distance to the host
galaxy of $8.4\,$Mpc, corresponding to a distance modulus of $\mu=29.6\,$mag
\citep{feldmeier97}.

\subsection{Reddening}
\label{sec:reddening}

The amount of extinction by dust has a significant impact on the estimate of the
total luminosity of the object and, therefore, the amount of synthesised
{\nifs} derived from models. The extinction toward SN~1994I has been a
controversial topic \citep{baron96,richmond96} and could not be firmly
established.  The presence of a strong, narrow \ion{Na}{i}~D line in the
spectra, as well as the colours, suggest that the object was significantly
reddened, but the exact amount is subject to large uncertainty.  While the
analysis of the \ion{Na}{i}~D line implies a very large extinction of
$\ebv \approx 1$ \citep{ho95}, spectral models by \citet{baron96} suggest a
value for $\ebv$ between $0.30$ and $0.45\,$mag. 

The amount of reddening affects the model spectra because the luminosity affects the
temperature and hence the properties of the spectrum (given a fixed distance
and density structure). The temperature structure determines the overall shape
of the spectrum and, via the ionisation balance, the strengths of characteristic
line features.  A test series of different values for $\ebv$ between $0.15$ and
$0.50\,$mag showed a preferred range between $0.25$ and $0.35\,$mag. For larger
reddening the ionisation tends to be too high to reproduce the characteristic
lines of singly ionised species \citep{millard99}. A value of $\ebv$ near the
lower end results in temperatures that shift the line ratios, in particular
those of \ion{Ca}{ii}, too much toward the transitions with low excitation
energies.  For our full series we chose $\ebv=0.30\,$mag, however note that the
uncertainty for this value may be as large as $\pm0.05\,$mag. For models with
larger reddening the composition required to fit a spectrum contains more
iron-group elements at the expense of lighter elements because a larger
reddening requires a larger temperature, shifting the spectrum to the blue.
This blue-shift has to be counterbalanced by stronger line blocking, which is
efficiently provided only by iron-group elements having a large number of lines.

In the following the properties of each model are briefly discussed.

\subsection{Model series for the photospheric epoch}
\begin{figure*}\centering
    \includegraphics[width=81mm]{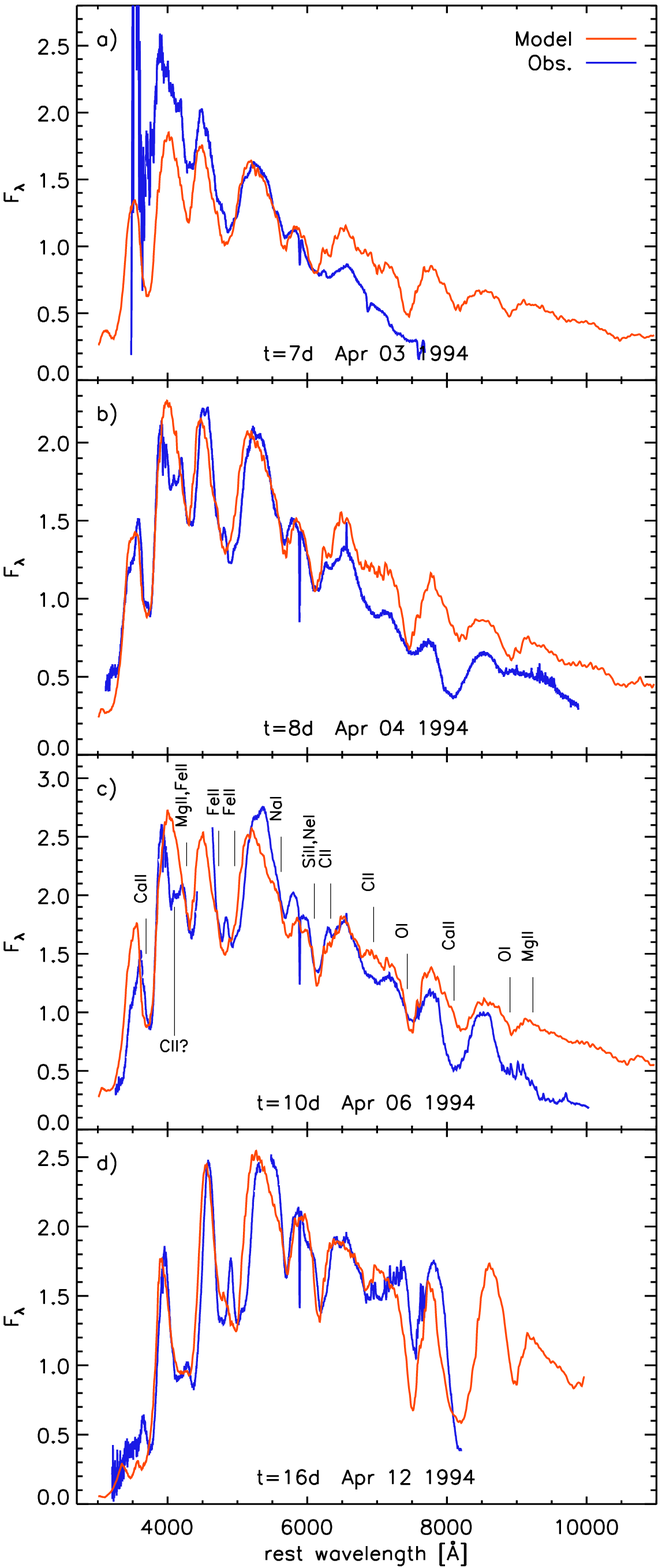}
    \includegraphics[width=81mm]{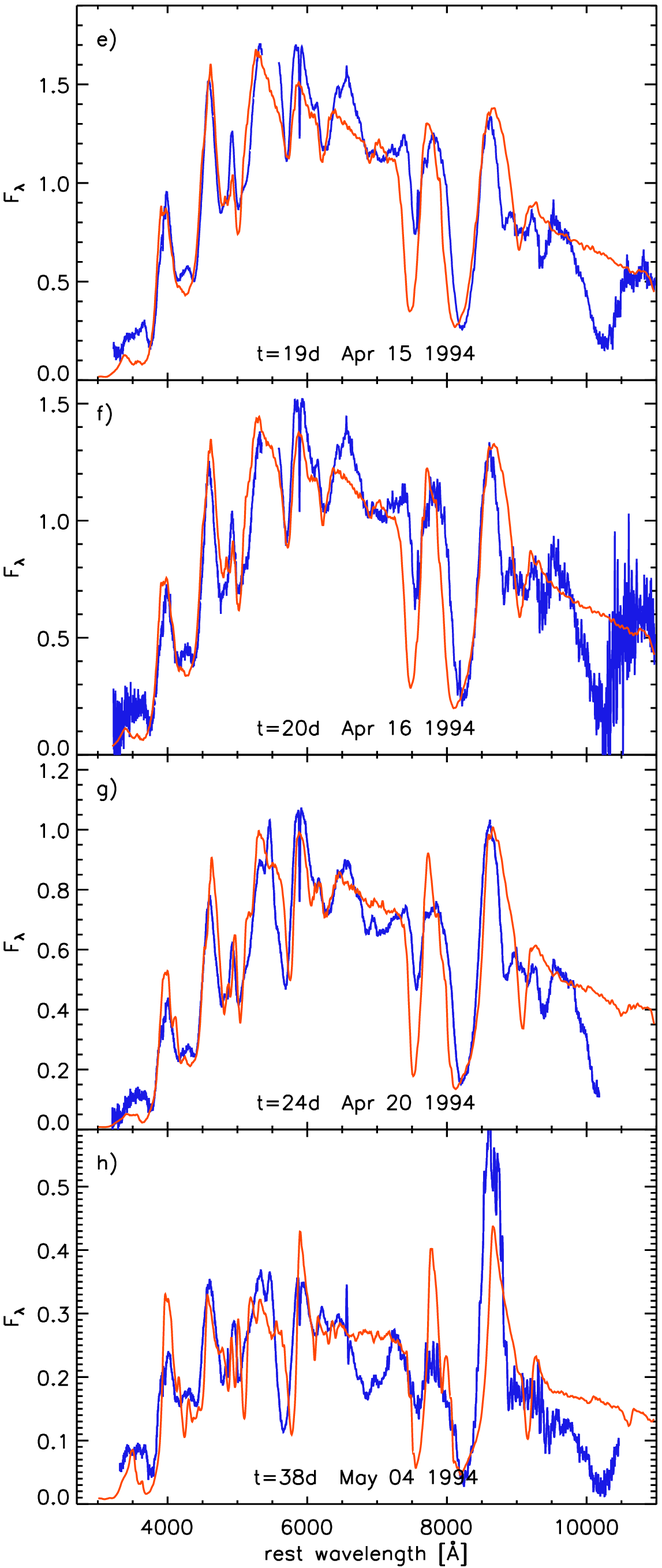}
  \caption{\label{fig:modelspec}Synthetic spectra of SN~1994I in
  comparison with the observed spectra. The observations are from
  \citet{filippenko95}, except for the spectrum of April 3, which was
  taken at the MMT (Schmidt et al. 1994, priv. comm.). For each spectrum
  the epoch in days after the assumed explosion date (March 27, 1994) is
  indicated. The flux $F_{\lambda}$ is given in units of
  $10^{-14}\,$erg\,cm$^{-2}$\,s$^{-1}$\,{\AA}$^{1}$. In panel c the ions that
  form the major contribution to the observed features are indicated. The
  identifications do not change significantly at different epochs.}
\end{figure*}

\subsubsection{$t=7\,$d, April 3, 1994}

For this spectrum the calibration against the corresponding photometry required
a correction of $0.17\,$mag to match the $R$ and $V$ bands reasonably well. 
However, with this correction the $U$ and $B$ bands are still too bright by
$0.67$ and $0.30\,$mag, respectively. This deviation is too large to apply a
single averaged value. Therefore we re-calibrated the spectrum to match only the
$V$ and $R$ bands, which represent the majority of the flux.  Keeping this
inconsistency in mind, we did not try to  match the observations exactly but
rather adjusted the model to obtain the basic features and reach agreement with
the $U$, $B$, $V$ and $R$ photometric points for this epoch.  Consequently, the fit does
not look too good at all wavelengths (see Fig.~\ref{fig:modelspec}a), but the main
features are reproduced, except for the \ion{O}{i} $\lam7774$ line, which is
significantly weaker in the observed spectrum. This line tends to be too strong
in all our models. It is a highly saturated line that is fairly insensitive to
changes in the oxygen abundance, within reasonable limits.

Apart from the inconsistency in the data, part of the problems to fit the
overall slope of the observed spectrum could also arise from the model
assumption of a thermal lower boundary which may not sufficiently represent the
physical conditions in the ejecta. We discuss this possibility in
Sec.~\ref{sec:disc-vph}.

\subsubsection{ $t=8\,$d, April 4, 1994}

The total luminosity required for this model is $L=1.89\times10^{42}\,${\ergs}. 
With a photospheric velocity of $v_{\rm ph}=11\,600\,${\kms} this results in a
temperature of the underlying blackbody of $T_{\rm BB}=9679\,$K. In the blue
and visible wavelength bands the fit is reasonable, while the slope of the
pseudo-continuum of the model does not match that observed in the redder parts
of the spectrum.  The spectrum shows various strong absorption features which
can be attributed to the ions indicated in Fig.\ref{fig:modelspec}c. 
Interestingly, the absorption at $\sim\!6100\,${\AA} is only partially produced
by \ion{Si}{ii}; it also contains a significant contribution from various
\ion{Ne}{i} lines, most important \ion{Ne}{i} $6402$. As before, the absorption
feature of the \ion{O}{i} $\lam7774$ line appears  too deep.

The model also does not reproduce the spectral slope of the observation redward
of about $6500\,${\AA}. Part of this problem may, as in the previous spectrum,
be due to a calibration problem in the data. The correction that was applied to
match the $V$ and $R$ photometry results in an offset of $0.14\,$mag in the $I$
band. Therefore, the $I$-band magnitude of the model is actually in closer
agreement with the photometry than the integrated magnitude of the spectrum in
that bandpass filter. Nevertheless, also for this case the possibility of a
physical effect that is not correctly represented in the model should be kept
in mind (see Sec.~\ref{sec:disc-vph}).


Another interesting feature of this spectrum is the absorption structure
between $4000\,${\AA} and $4200\,${\AA} which \citet{millard99} attribute to
\ion{Sc}{ii}.  In our models  we can reproduce that feature using significant
amounts of Sc, but this also causes a variety of other Sc lines to show up
elsewhere in the spectrum.  In the later spectra the observed absorption seems
to grow into a distinct feature which is mostly caused by \ion{Ti}{ii} and
\ion{Fe}{ii} lines, while there is no clear indication for Sc lines anymore.
Therefore we suggest that even in the early spectrum the main components of
this feature are Fe lines that are not correctly reproduced by our model. In
addition, other ions such as \ion{C}{ii} and \ion{Si}{ii} contribute to this
feature to some extent.

\subsubsection{$t=10\,$d, April 6, 1994}

This spectrum (Fig.~\ref{fig:modelspec}c) represents the observation closest to
the $B$-band maximum, which occurred on April 7 \citep{richmond96}. In this
model, the continuum slope is already better reproduced than in the first
spectra although there is still an offset redward of $\sim\!7000\,${\AA}.  We
used a photospheric velocity of $v_{\rm ph}=10\,800\,${\kms}. The total
luminosity increased to $L=2.31\times 10^{42}\,${\ergs}, and the temperature of
the incident radiation was $T_{\rm BB}=9230\,$K. With the exception of the
``Sc'' feature and red continuum slope, the fit is reasonably good. The
observation starts to show the separation of the \ion{Fe}{ii} feature at
$4900\,${\AA} which is reproduced by the model at this epoch. Also, the depth
of the \ion{Ca}{ii} infrared lines is not sufficient in the model while the
\ion{Ca}{ii} H\&K absorption is fitted well.

\subsubsection{$t=16\,$d, April 12, 1994}

This spectrum (Fig.~\ref{fig:modelspec}d), $16\,$d after explosion
(corresponding to $5\,$d after $B$-maximum), exhibits clear differences
compared to the pre-maximum spectra, and it is significantly redder.  The
parameters that have been used for this fit are $L= 2.10\times10^{42}\,${\ergs}
for the total luminosity and a photospheric velocity of $v_{\rm ph}=8800\,${\kms}. 
The resulting temperature at the inner boundary was $T_{\rm BB}=7690\,$K.  The
double structure between $4500\,$ and $5000\,${\AA} is more pronounced, and our
model mostly fails to reproduce it.  The feature is mainly formed by
\ion{Fe}{ii} lines that are all part of the same multiplet (\ion{Fe}{ii}
$\lam\lam4923, 5018, 5169$). The fact that we see the lines separated in the
observation indicates that either the density or at least the ionisation
fraction of \ion{Fe}{ii} must have a steep gradient in the velocity region just
above the photosphere, which is not reproduced by our models of the earlier
epochs. At later epochs the models resolve this structure better.

Apart from that, most of the features in the model are systematically
blue-shifted compared to the observation.  Even with a drastic change of
parameters this problem remained, suggesting that the density structure is too
flat within the velocity range where most lines form at this epoch.

\subsubsection{$t=19\,$d, April 15, 1994}

The luminosity used for this model (see Fig.~\ref{fig:modelspec}e) decreased
further to $L=1.46\times10^{42}\,${\ergs} at a photosphere located at
$v_{\rm ph}=7200\,${\kms} with a temperature of $T_{\rm BB}=6999\,$K.  In this
spectrum the double feature of Fe lines below $5000\,${\AA} also starts to show
in the synthetic spectrum.

Starting from this epoch the slope of the near-IR flux is well reproduced by
the model. This may be surprising because at later epochs the approximation of
a photosphere becomes less accurate. Given the compactness of model CO21, it
may be that the photospheric approximation holds reasonably well even at epochs
well after maximum (see \S\ref{sec:disc-vph}).

The model does not provide clear evidence for the origin of the deep absorption
observed at $10\,500\,${\AA}.  In previous work \citep{filippenko95,
clocchiatti96, millard99, baron99}, various suggestions for the origin of this
feature have been discussed, but no conclusive answer could be reached. We will
investigate this feature in \S\ref{sec:irfeature}

\subsubsection{$t=20\,$d, April 16, 1994}

This spectrum (Fig.~\ref{fig:modelspec}f) is just a day after the previous one
and therefore fairly similar. A model with $L=1.30\times10^{42}\,${\ergs},
$v_{\rm ph}= 6800\,${\kms}, and $T_{\rm BB}=6774\,$K reproduces most features
quite well, except that some of the peaks tend to be somewhat low.

The narrow peak at $6560\,${\AA}, which is not reproduced by our model, could
to some extent be due to emission of H$\alpha$ by a surrounding shell. To test
this possibility, however, is beyond the scope of this work.

\subsubsection{$t=24\,$d, April 20, 1994}

The $t=24\,$d epoch represents the last of the early spectral series.  Overall
this spectrum is still similar to the one taken $4\,$ days earlier, except that
the \ion{Fe}{ii} and \ion{Na}{i} lines become deeper. For the model in
Fig.~\ref{fig:modelspec}g we used $L= 9.18\times10^{41}\,${\ergs} and a
photospheric velocity of $v_{\rm ph}= 5400\,${\kms}. The temperature of the
photosphere was derived to be $T_{\rm BB}=6383\,$K.

\subsubsection{$t=38\,$d, May 4, 1994}

This spectrum, $38$ days after explosion, is already quite late to be treated
with a photospheric code because it may contain a significant fraction of
nebular-line emission mixed into the emission components of the P-Cygni
profiles.  Therefore, it is not surprising that some of the emission features
cannot be reproduced by this approach (e.g., the \ion{Ca}{ii} IR triplet). 
Nevertheless, the fit is reasonable (see Fig.~\ref{fig:modelspec}h).  The
luminosity used was $L= 3.49\times10^{41}\,${\ergs}; the photosphere was set to
$v_{\rm ph}= 2100\,${\kms}. These values are not strongly constrained because a
clear definition of a photosphere is not possible at such late epochs. 
Rather, the velocity sets the radius which is required to obtain the correct
temperature structure to reproduce the correct ionisation ratios given the
chosen luminosity. The temperature at the photosphere was $T_{\rm BB}=6129\,$K.

\subsection{Nebular Spectrum, $\bmath{t=159}\,$d}\label{sec:neb-model}

As is typical for {\sneic}, the nebular phase develops early because of the
small ejecta mass.  To test the inner part of the ejecta we also modelled the
nebular spectrum taken on September 2 using a non-LTE nebular code
\citep{mazzali01a} which assumes a constant density, one-zone model inside a
given velocity.

\begin{figure}
 \includegraphics[width=8.4cm]{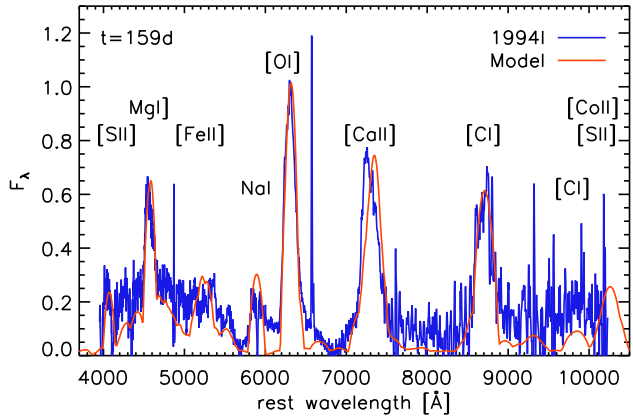}
  \caption{\label{fig:nebsp} Nebular model for the nebular spectrum
    $159\,$d after explosion compared to the observation assuming
    a reddening of $\ebv=0.30\,$mag.  $F_{\lambda}$ is in units of
  $10^{-15}\,$erg\,cm$^{-2}$\,s$^{-1}$\,{\AA}$^{1}$}
\end{figure}

The spectrum has an epoch of $159\,$d after the proposed explosion date.   It
was corrected based on the photometric data of \citet{richmond96} but in the
blue part the correction is somewhat uncertain because of the presence of an
underlying galaxy continuum.  However, this uncertainty does not significantly
affect the conclusions in this section.

For the model shown in Fig.~\ref{fig:nebsp} we assumed an expansion velocity of
$v_{\rm exp}=5500\,${\kms} at an epoch of $t=159\,$d. The model has a mass of
$M=0.43{\msun}$ below this velocity. To reproduce the correct line strengths
using the same distance and reddening values as for the photospheric spectral
series, $0.07\msun$ of {\nifs} are required. This generates a luminosity of the
nebula of $L=2.22\times10^{40}\,${\ergs} with a faction of $\sim\!10\,$per cent
of the $\gamma$-rays depositing their energy in the gas.  The remaining
composition needed to reproduce the spectrum includes $0.22\msun$ oxygen,
$0.09\,\msun$ carbon, $0.02\msun$ silicon, and $0.02\msun$ sulphur.

\begin{table}
  \caption{\label{table:neb-red}Nebular models for different values of reddening.}
  \begin{tabular}{@{}lccc}
    \hline
    $E(\bv)$ [mag]   & $0.15$ & $0.30$ & $0.45$  \\
    \hline
    $M_{\rm total}$ [$M_{\sun}$]   & $0.36$ & $0.43$ & $0.54$ \\
    $M(\nifs)$ [$M_{\sun}$]      & $0.05$ & $0.07$ & $0.09$ \\
    $M({\rm O})$ [$M_{\sun}$]   & $0.18$ & $0.22$ & $0.28$ \\
    $L_{\rm dep}$ [$10^{40}\,${\ergs}]   & $1.42$ & $2.22$ & $3.31$ \\
    \hline
  \end{tabular}
\end{table}

To test the impact of reddening we calculated a series of models with different
values for $\ebv$: $0.15$, $0.30$, and $0.45\,$mag.  Within this range of $\ebv$
it is possible to find a set of parameters that fit the observed spectrum well,
but the masses of the various models are different. The resulting parameters
for these models are summarised in Table~\ref{table:neb-red}.

The mass enclosed within $5500\,\kms$ is larger than the corresponding mass of
CO21, which is only $0.1\msun$. Also, the line profiles do not suggest the
presence of a mass cut at a velocity of $\sim\!2000\,\kms$ but indicate the
presence of material at lower velocities.

\section{Discussion}\label{sec:discussion}

\subsection{Composition and derived magnitudes}

\begin{figure}
 \includegraphics[width=8.4cm]{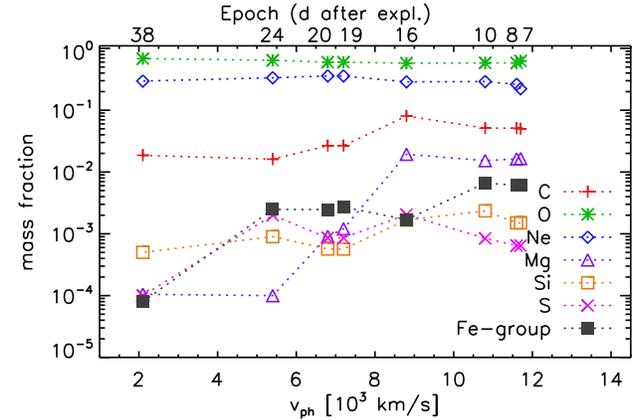}
  \caption{\label{fig:comp}Mass fraction of the various elements versus the velocity of
  the photosphere ($v_{\rm ph}$) that was used for the respective model.
  At each epoch the models are calculated assuming a homogeneous
  distribution of the elements.}
\end{figure}
Fig.~\ref{fig:comp} shows a plot of the mass fraction for some elements used
for the respective models versus the photospheric velocity, $v_{\rm ph}$. Note
that we assume a homogeneous composition for each model; hence, this plot will
only roughly represent a slice of the ejecta composition. In particular, for
features that form significantly above the photosphere, the velocity may not be
representative. Also, for elements that do not contribute any distinct line
features, the abundance is not strongly constrained.

An interesting aspect of this plot is that the abundance of iron-group elements
(Ti, Cr, Fe, Co, and Ni) {\em decreases} toward lower photospheric velocities.
This is a feature which is seen in hypernovae that exhibit strong jet-like
asymmetries. The fact that we find a higher concentration of heavier elements
in the outer part of SN~1994I may suggest that the explosion of this object
also showed asymmetry to some extent, although the nebular spectra do not show
a strong indication for that.

In comparison to the original CO21 model, we get a significantly larger mass
fraction of oxygen at low velocities at the expense of iron-group elements. The
high abundance of iron-group elements at velocities above $\sim\!7000\,\kms$ is
also not predicted by CO21 even if we assume some contribution of primordial
material with solar composition added into the outer part of the explosion
model. The abundance of Si is significantly lower than predicted by CO21, while
C and Ne seem to be more abundant.  On the basis of the spectral  models it is
not possible to put strong constraints on the detailed composition structure;
in particular, for elements that do not have many spectral lines, the error
bars may be quite large.  Nevertheless, there is strong indication that the
composition is more strongly mixed than in the theoretical ejecta model.

\begin{figure}
 \includegraphics[width=8.4cm]{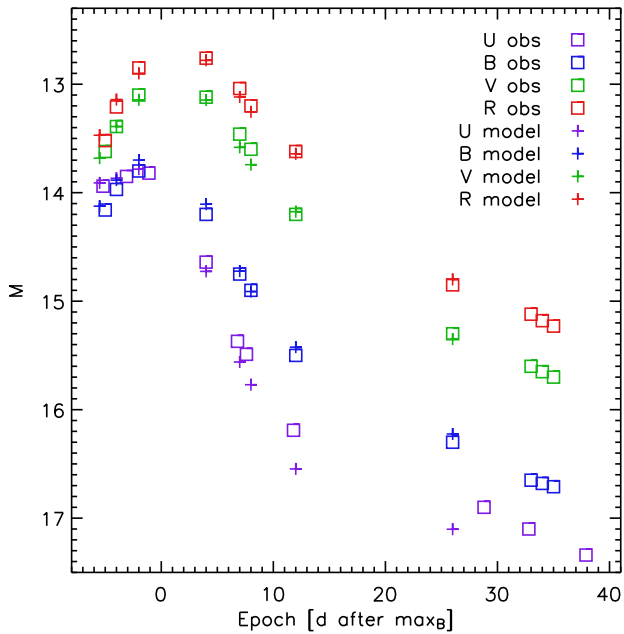}
  \caption{\label{fig:mags}Comparison of observed (squares) and model (crosses) magnitudes
  in different wavelength bands. The observed magnitudes for $B$,
  $V$, and $R$ have
  been obtained by interpolating the photometry points from
  \citet{richmond96} to the observation dates of the spectra.
  The $U$-band points are from Schmidt, et al. (1994, priv. comm.).}
\end{figure}

Fig.~\ref{fig:mags} shows the magnitudes derived from the models in comparison
to the ``observed'' magnitudes. The data points for $B$, $V$, and $R$ have been
obtained by interpolating the original points of \citet{richmond96} to the
epochs of the corresponding spectral observations.  The $U$-band points are
taken from Schmidt, et al.  (1994, priv. comm.) and are shown at the original
epochs because there is insufficient spectral coverage of this wavelength band. 
Since the models resemble the observed spectral shapes reasonably well, the
magnitudes from the models are overall in good agreement with the observed
ones. Only at the later epochs the $U$-band magnitudes are underestimated by the
models.

\subsection{The location of the ``photosphere''}
\label{sec:disc-vph}

\begin{figure}
 \includegraphics[width=8.4cm]{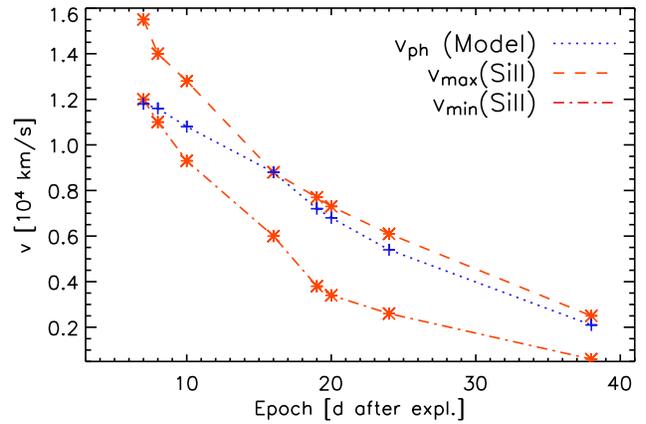}
  \caption{\label{fig:t-v}Photospheric velocities of the models as
  a function of epoch in comparison to the velocities measured from the
  ``\ion{Si}{ii}'' feature at $6100\,${\AA} (see text).}
\end{figure}
Fig.~\ref{fig:t-v} shows the photospheric velocities ($v_{\rm ph}$) derived
from the models as a function of epoch.  Also indicated are the measurements of
the line velocity from the observed spectra {\em assuming} that the $\lam6100$
feature is due to \ion{Si}{ii}$\,6347,6372$. Our models suggest that there may
be a significant contribution from \ion{Ne}{i} $6402$. This would result in an
underestimate of the ``real'' velocity because the \ion{Ne}{i} line has a
slightly redder rest wavelength. The upper and lower limits correspond to
different assumptions on the blend of lines that makes this feature. The lower
dotted curve corresponds to a fit to the feature with a single Gaussian
profile, while the upper dotted curve represents the estimate from a fit with
two Gaussian profiles assuming that the bluer one is \ion{Si}{ii}.  At the
early epochs the photospheric velocities of the models clearly exhibit a
different trend. Therefore, the absolute velocities deduced in this way may not
be a good representation of the photospheric velocities and should not be
over-interpreted.  

Because the photospheric velocity of the models are primarily constrained by the
temperature that is needed to match the shape of the spectrum and the degree of
ionisation, this different behaviour of the \ion{Si}{ii} velocities may result
from the failure of the approximation to describe the shape of the early-time
spectra with a thermal blackbody continuum. The mismatch between the observed
and model red continua give another indication that this description is not
appropriate for those epochs. Generally, in hydrogen-deficient supernovae the
assumption of a thermal continuum photosphere is questionable because the
physical conditions present in the ejecta do not result in sufficient true
continuum opacity to achieve thermalization \citep[e.g.,][Sauer et~al, in
prep.]{sauer05}.  Imposing this thermal continuum in the models generally leads
to an offset of the model flux in the red and infrared wavelength bands. 

This problem is less severe in {\sneic} than in {\sneia} because the
composition by mass fraction  is dominated by the lighter species (oxygen and
neon) while in {\sneia} the iron-group elements make the largest contribution. 
This means that even for the same mass density, the number density in {\sneic}
is larger by a factor of up to $2$. Since the true processes (bound-free and
free-free transitions) that are needed to thermalize photons scale with the
number density of the ions and free electrons, the processes are more effective
in {\sneic}. In addition, the stronger influence of Thomson scattering (which
is also proportional to $n_{e}$) will result in a larger number of scattering
events for each photon before it can escape the ejecta.  This effectively
increases the volume in which the photons have a chance of being thermalized. 
This may explain why the spectral shape of SN1c spectra can be described
fairly well with a thermal continuum to later times whereas in SNe~Ia, the
approximation to assume a blackbody continuum at the photosphere becomes
increasingly worse for later times.

The early-time spectra of SN~1994I  are very blue suggesting that they may have
a stronger component of non-thermalized radiation than the later ones. This may
be due to a steep gradient in the density in the outer parts of the ejecta. If
a substantial quantity of $\gamma$-ray photons is generated in the outer,
thinner layers of the ejecta, the effective path lengths of the photons may
become too short to allow the photons to be thermalized.  A prerequisite for
the presence of a non-thermalized radiation field at early times is, however, a
significant amount of radioactive {\nifs} located in the outermost layers of
the ejecta.  This outer layer of Ni would not strongly contribute to the later
light curve, as a large fraction of the $\gamma$-ray photons would escape
without depositing their energy in the ejecta.  Some {\nifs} is indeed needed
in the outer part of the ejecta to reproduce the early light curve rise (see
\S\ref{sec:lc-model})

For a more consistent investigation of this phenomenon a better
description that includes the non-thermal formation of the pseudo-continuum is
needed.

\subsection{The infrared feature at 10\,300\,{\AA}}\label{sec:irfeature}

\begin{figure}
  \includegraphics[width=8.4cm]{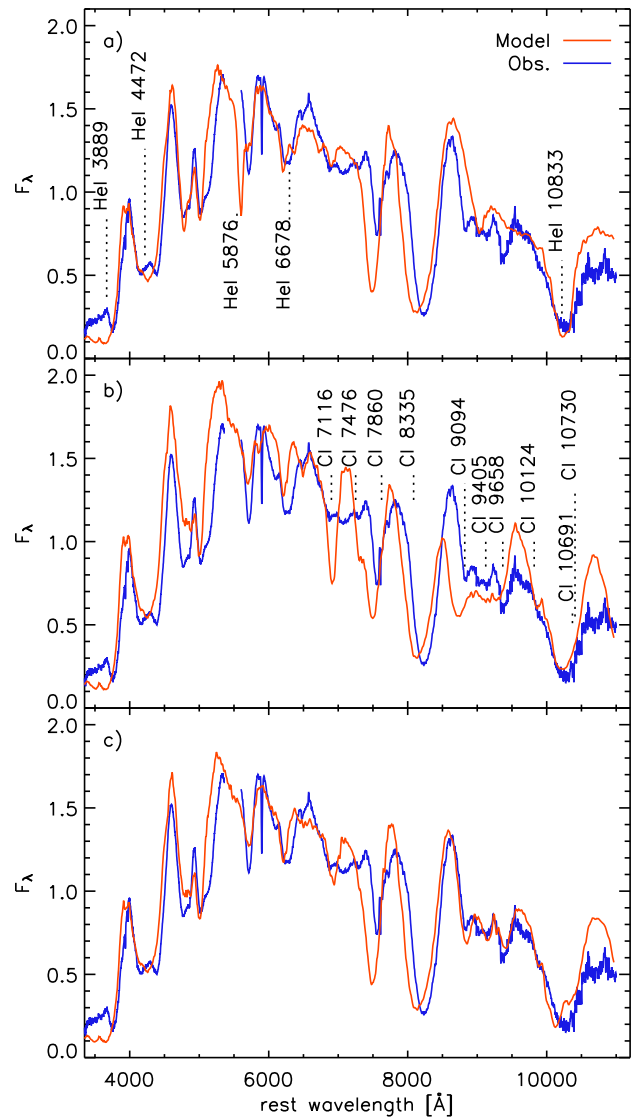}
  \caption{\label{fig:irfeature}
  Models for the $t=19\,$d spectrum. In the upper panel (a) the strength of the
  helium lines was artificially enhanced above $v\approx16\,000\,\kms$ by
  factors between $10^{8}$ and $10^{12}$  to mimic non-thermal excitation in an
  outer helium-rich shell (assuming a  homogeneous mass fraction of $10.5$ per
  cent for helium). It is not possible to reproduce the broad absorption
  feature at $10\,300\,${\AA} in this way without also increasing the strength
  of the \ion{He}{i} $\lam 5876$ line to a level that is not observed. The
  labels in Panel~a indicate the positions of the strongest \ion{He}{i} lines
  at a velocity of $v_{\rm He}=17\,000\,$\kms. The middle panel (b) shows a
  model obtained by increasing the optical depth in the \ion{C}{i} line by
  factors of $10^{3}$ to $2\times10^{3}$ above $14\,000\,${\kms} to fit the
  $10\,300\,${\AA} absorption. The mass fraction of carbon was $21.7$ per cent. 
  This model shows a variety of \ion{C}{i} lines in other places that are not
  observed. The positions of the \ion{C}{i} lines correspond to a velocity of
  $v_{\rm C}=9000\,\kms$. The lower panel (c) shows a model with both
  \ion{He}{i} and \ion{C}{i} enhanced. The enhancement factor for \ion{He}{i}
  above $16\,600\,${\kms} reaches from $10^{9}$ to $10^{12}$, for \ion{C}{i} we
  applied a factor of $50$ in the entire ejecta.  This leads to an acceptable
  fit of not only the IR-absorption but also various other lines at
  $\sim\!9000\,${\AA}, which are not reproduced by the original model
  (Fig.~\ref{fig:modelspec}e).  The flux is given in units of $10^{-14}\,$erg\,cm$^{-2}$\,s$^{-1}$\,{\AA}$^{1}$.}
\end{figure}
Spectra that extend into the near infrared (Fig.~\ref{fig:modelspec}e-h) show a
distinct and broad absorption feature around $10\,300\,${\AA}.  Originally,
this feature was attributed to \ion{He}{i} $\lam10\,830$
\citep{filippenko95,clocchiatti96}.  However, as pointed out by
\citet{baron99}, this identification would require a detection of at least
\ion{He}{i} $\lam5876$ at similar velocities ($\sim\!15\,000\,${\kms}) in the
observed spectrum.  To model the \ion{He}{i} lines
correctly it is necessary to consider the non-thermal excitations by fast
electrons caused by Compton scattering of $\gamma$-photons from the $^{56}$Co
decay. Without those processes the excitation of \ion{He}{i} at
the temperatures present in the ejecta is too low to cause any significant
absorption \citep{lucy91}. 

Since our code does not treat non-thermal processes we cannot address this
problem in a self-consistent way. To nevertheless test the possibility that
\ion{He}{i} causes the $10\,300\,${\AA} absorption we enhanced the optical
depth in the \ion{He}{i} lines introducing by hand a ``non-thermal factor''
that mimics the effects of departure from LTE caused by the fast electrons
produced by the deposition of $\gamma$-photons
\citep[see][]{harkness87,tominaga05}. We allowed this factor to vary with      
velocity to obtain a good fit of the $10\,300\,${\kms} absorption. Although
some lines arise from the singlet system  and others from the triplet system,
such as $\lam6678$, we applied a single enhancement factor to all \ion{He}{i}
lines. The studies of           \citet{lucy91} and \citet{mazzali98a} find that
the departure from LTE is   comparable for both states.  We chose the observed
spectrum of $t=19\,$d (cf.   Fig.~\ref{fig:modelspec}e) for this study because
that is the earliest observation that extends far enough to the IR to show this
feature. In the first model shown in Fig.~\ref{fig:irfeature}a we enhanced the
optical depth in all \ion{He}{i} by a factor that varied from $10^{8}$ at
$16\,000\,${\kms} to $10^{12}$ in the outermost shells above $30\,000\,${\kms}.
The abundance of helium was uniformly set to $10.5$ per cent by mass.  Most of
the strong \ion{He}{i} lines that also show up in the model spectrum blend in
with other features in the spectrum, however the model inevitably shows a clear
absorption at $\sim\!5500\,${\AA}.  $\lam10\,833$ is the strongest
{\ion{He}{i}} line in our model and thus may show up before any other lines can
be clearly identified.  Unfortunately, the observed spectra that cover the
infrared wavelengths have a gap in the region where the \ion{He}{i} $\lam5876$
lines would be expected. The observed spectra of the later epochs that do cover
this wavelength range show a small absorption dip in the peak around $5450\,${\AA},
which could be identified with the \ion{He}{i} $\lam5876$ lines, but the
feature is not clearly present in the early epoch spectra. Apart from this, the
observed absorption is much weaker than what the model spectrum would suggest
for this line.  Also, the corresponding velocity for \ion{He}{i} would be
$26\,000\,${\kms}, which is inconsistent with the identification of \ion{He}{i}
$\lam10\,833$ for the $\lam10\,300$ absorption.  Of course, the method adds
more free parameters and therefore the abundance of {\ion{He}{i}} cannot be
constrained but depends on the choice of the enhancement factor (for a given
density of the explosion model).  The very large enhancement factors needed to
obtain a reasonable fit, however, indicate that an unreasonably large amount of
several solar masses of helium would have to be present at the high velocities
to obtain this absorption considering that to account for the non-thermal
excitation alone should not require a factor of more than a few $10^{5}$
\citep{lucy91,mazzali98a}.

\citet{millard99} argue from their highly parametrized line fits with SYNOW
that the $10\,300\,${\AA} feature could originate from a blend of \ion{He}{i},
\ion{Si}{i}, and \ion{C}{i} lines.  In our models of these epochs the ionisation
equilibrium is generally such that \ion{Si}{ii} is more abundant than
\ion{Si}{i} by several orders of magnitude.  Therefore, the mass in Si needed
to obtain considerable absorption in \ion{Si}{i} is very high and leads to a
significant overestimate of the \ion{Si}{ii} lines in the spectra.  The
strongest line of {\ion{Si}{i}} in the observed wavelength range is
{\ion{Si}{i}} $\lam10\,790$. If we assume that this line is primarily
responsible for the observed IR absorption, it follows that the velocity of
this {\ion{Si}{i}} layer should be around $15\,500\,${\kms}. In that case one
would also expect the presence of absorption features by the much stronger
{\ion{Si}{ii}} $\lam\lam6241,6349$ lines but none of the spectra suggest the
presence of such a high-velocity silicon layer.  Given those indications, we
regard the identification with {\ion{Si}{i}} as unlikely.

The presence of carbon at velocities below $11\,000\,${\kms} at day $19$ is
consistent with the earlier spectra at $t=8\,$d and $10\,$d: The small features
in the peak around $4100\,${\AA} in those early spectra (see line label in
Fig.~\ref{fig:modelspec}c) can be partially attributed to \ion{C}{ii} $\lam4267$
at a velocity around $11\,000\,${\kms}, although our models do not reproduce
the details of the spectrum in that region  The smaller absorption features
between the \ion{Ca}{ii} IR-triplet and the deep absorption at $10\,300\,${\AA}
in the later spectra can be identified with {\ion{C}{i}} lines. These lines are
indicated in Fig.~\ref{fig:irfeature}b, which shows a model in which the optical
depth in \ion{C}{i} was enhanced above $14\,000\,\kms$ by factors between $10^{3}$ and
$2\times10^{3}$ to fit the $10\,300\,${\AA} absorption. The mass
fraction of carbon in this model is $21.7$ per cent. The positions of those
lines correspond to a velocity of around $9500\,${\kms}, which also agrees with
the velocity of the {\ion{O}{i}} lines near $7500\,${\AA}. The enhancement
factors needed for \ion{C}{i} are significantly lower than for helium. In
addition, the observed spectra of earlier epochs are consistent with the
presence of {\ion{C}{ii}} lines at similar velocities. Nevertheless, the model
shown in Fig.~\ref{fig:irfeature}b does also not provide a reasonable fit to
other spectral regions. That suggests that carbon alone may also not be the
cause of the $10\,300\,${\AA} absorption.

The lower panel in Fig.~\ref{fig:irfeature} shows a model with different
factors mimicking non-thermal excitations in both \ion{He}{i} and \ion{C}{i}. 
This combination gives a good fit not only to the IR-absorption but also to the
\ion{C}{i}-lines shown in Fig.~\ref{fig:irfeature}b. In addition, the amount of
\ion{He}{i} needed is low enough not to cause strong absorptions elsewhere in
the spectrum. For this model the optical depth in \ion{He}{i} was enhanced by
factors between $10^{9}$ to $10^{12}$ above $16\,600\,\kms$. For \ion{C}{i} a
uniform factor of $50$ was applied in the entire ejecta.  The relatively small
enhancement-factors needed for \ion{C}{i} and the fact that one would not
expect that non-thermal effects are significant for this ion, suggest that a
slightly different composition and ionization structure at larger velocities may
be responsible for the formation of the \ion{C}{i} lines.

For a more conclusive determination the effect of varying composition as well
as a more physical description of the non-thermal excitations has to be
considered.  Also, as pointed out by e.g.  \citet{przybilla05}, in full non-LTE
calculations of early type stars the strength of the IR-{\ion{He}{i}} line
sensitively depends on the correct treatment of line blocking as well as the
details of the chosen model atom.  Therefore, fully consistent non-LTE models
will be necessary to model the formation of this line.

\subsection{Oxygen abundance and total mass}\label{sec:oxygen}

Common to all synthetic fits of the photospheric epochs presented in
\S\ref{sec:ph-models}is  the fact that the predicted \ion{O}{i} $\lam7774$ line
that is too strong. This line is highly saturated and does not depend
sensitively on the abundance assumed for oxygen. Therefore, one has to adopt a
very low oxygen abundance to fit the observed depth of the \ion{O}{i} line.  In
the $t=19\,$d spectrum, for example, the oxygen mass fraction has to be reduced
to $\sim\!4$ per cent.  At the same time, however, a significant oxygen mass of
$0.22\msun$ is needed in the innermost region to reproduce the nebular spectrum
(see \S\ref{sec:neb-model}).

Another possibility is that the total oxygen mass fraction is indeed larger,
but that, owing to non-thermal ionisation processes the ionisation of
{\ion{O}{i}} is actually higher than derived by our models.  {\ion{O}{ii}}
does not have many lines from low-lying levels in the observed wavelength
bands.  The presence of unexplained absorption dips in the spectra around
$7000\,${\AA} could hint at forbidden {\ion{O}{ii}} lines, which have also been
suggested by \citet{millard99}.  Non-thermal ionisation can be induced by some
amount of mixing of {\nifs}-rich material into the oxygen layers. 
The larger amount of oxygen in the inner region seen in the nebular phase
supports this scenario.

The obvious contradiction in this picture, however, lies in the identification
of {\ion{C}{i}} lines in the infrared as discussed in the previous section. If
{\ion{C}{i}} is seen at velocities similar to that of the {\ion{O}{i}} line, it
is not clear why non-thermal ionisation processes should affect oxygen but not
carbon.

\subsection{Bolometric light curve}\label{sec:lc-bol}

\begin{figure}
  \includegraphics[width=8.4cm]{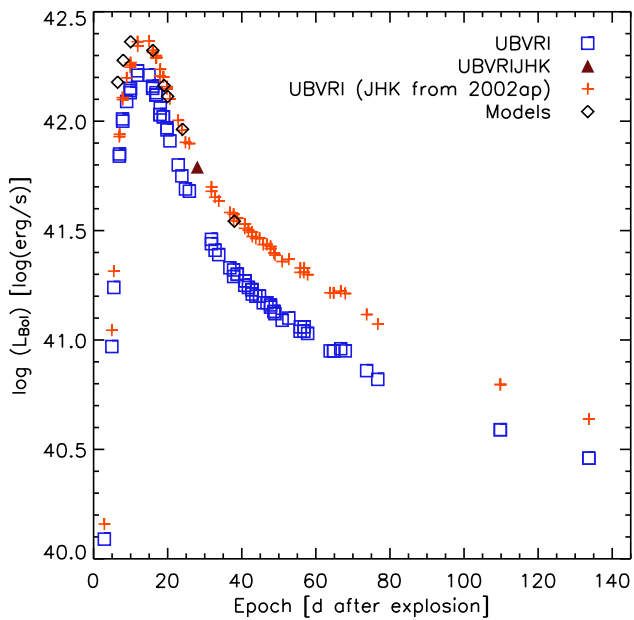}
  \caption{\label{fig:lcbol}Bolometric light curves for SN~1994I. The
  squares denote the bolometric light curve obtained from {\ubvri}
  observations of SN~1994I without correction for the unobserved {\jhk}
  bands. The filled triangle indicates the point for which {\jhk} is 
  also known.
  The crosses are the data points obtained by correcting the SN~1994I points
  for the unobserved {\jhk} bands by assuming the same contribution as in
  SN~2002ap. The results of the spectral models are shown as diamonds.}
\end{figure}

The bolometric light curve is one of the common ways to constrain a theoretical
model. Ideally,  to obtain a bolometric light curve from observed data one uses
a set of well-calibrated spectra from UV through IR wavelengths.  The available
data for SN~1994I does not cover the {\jhk} bands in the infrared.  We computed
the bolometric light curve of SN~1994I using the photometric data in  {\ubvri}
(\citealt{richmond96} and B.~Schmidt, priv. comm.). This {\ubvri} integrated
light curve from SN~1994I data (open squares in Fig.~\ref{fig:lcbol}) alone
represents a lower limit to the bolometric light curve since the IR
contribution is not taken into account. 
In addition, we derived an upper limit estimate by adding the IR-contribution
of SN~2002ap \citep{yoshii03}, which is the most similar object for which
infrared data have been published. For this addition we assume that the
relative contribution of the infrared bands in SN~1994I was the same as for
SN~2002ap after the data points of SN~2002ap have been rescaled to the
appropriate extinction.  The upper limits are shown as crosses in
Fig.~\ref{fig:lcbol}.  For SN~1994I additionally one point in each of $J$ and
$H$ and two in the $K$ band are available from \citet{rudy94}. This allows us
to compute a single bolometric point for the epoch $t=28\,$d after explosion
which is indicated by the filled triangle in Fig.~\ref{fig:lcbol}.  This single
point agrees nicely with the upper limit light curve which justifies the choice
of the SN~2002ap data to derive the upper limit light curve for SN~1994I. It
also shows that the IR contribution to the bolometric flux is not negligible. 
Generally, for a {\snic} the contribution of the IR-wavelength bands ranges
from about $20$ per cent before $B$ maximum, to roughly $40$ per cent near $K$
maximum. In the late phase the IR-bands contribute about $30$ per cent
\citep{yoshii03,patat01}.   

To compute the bolometric light curves, we converted the magnitudes into fluxes
using the standard {\ubvri} Johnson-Cousin system \citep{buser78,bessell83}. 
Using $V$ epochs for reference, we then interpolated to these epochs the flux
in other bands. The monochromatic fluxes were used as low-resolution spectra
and transformed to luminosities adopting a distance modulus of $m-M=29.6\,$mag,
an extinction of $\ebv= 0.3\,$mag (see \S\ref{sec:reddening}), and the
extinction law of \citet{savage79}.  

The luminosities used in our spectral models are shown as diamonds in
Fig.~\ref{fig:lcbol}. They are in good overall agreement with the estimate from
the observations corrected for the infrared contribution.

\section{Light curve models}
\label{sec:lc-model}

\begin{figure}
  \includegraphics[width=8.4cm]{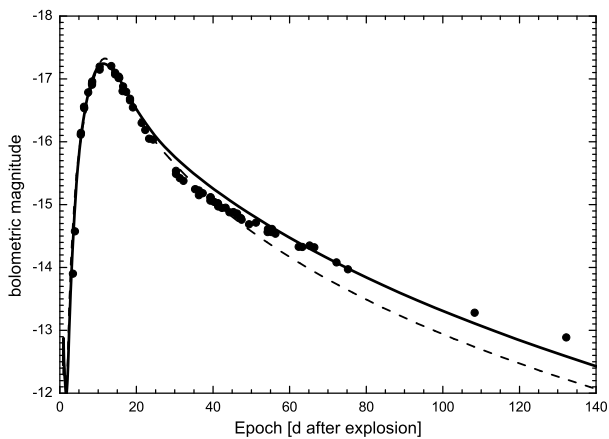}
  \caption{\label{fig:lcmodel}Synthetic light curve for the modified CO21
  ejecta model (solid line) in comparison with the light curve obtained
  from the original CO21 model (dashed line). The filled circles show the
  bolometric light curve from the observations (see \S\ref{sec:lc-bol}).}
\end{figure}

Fig.~\ref{fig:lcmodel} shows the synthetic light curve of the ejecta model that
we built, as suggested by the nebular-spectrum model, by adding a low-velocity
dense core to the density structure of CO21 ({\em solid line}), compared with
the light curve of the original CO21 ({\em dashed line}) and with the
bolometric light curve of SN~1994I ({\em circles}).  We used the
one-dimensional supernova light curve synthesis code developed by
\citet{iwamoto00}. The code solves the energy and momentum equations of the
radiation plus gas in the co-moving frame and the gray $\gamma$-ray transfer
simultaneously; see \citet{mazzali02}, \citet{deng05}, and \citet{mazzali06}
for our approximations on optical line opacities and Eddington factors. The
$^{56}$Ni mass and its distribution were adjusted to optimise the fit. For the
relative abundances of other elements, we used the results of our spectral
modelling.

Our ejecta model has a total mass of $\sim\!1.2 \msun$, of which
$\sim\!0.4\msun$ is distributed below $5500\,\kms$ as required in the nebular
spectrum modelling (see \S~2.3). For comparison, CO21 has in total
$\sim\!0.9 \msun$ with only $\sim\!0.1 \msun$ below $5500\,\kms$. The total
kinetic energy, $\sim\!1\times 10^{51}\,$erg, is barely changed by the addition
of the low-velocity dense core. The best light curve fitting $^{56}$Ni mass is
$\sim\!0.07\msun$ for both ejecta models, although the distribution of $^{56}$Ni
is different. For CO21, we filled its low-velocity region with as much $^{56}$Ni
as possible without spoiling the fit to the peak to increase the energy
deposition rate at late times. This results in $\sim\!0.04$--$0.05\msun$ of
$^{56}$Ni in the inner $\sim\!0.1\msun$ ejecta at $v<5500\,\kms$. For our
ejecta model, which has a core as massive as $\sim\!0.4 \msun$, a $^{56}$Ni
abundance of $\sim\!10$ per cent in that core is enough to reproduce the slow
decline of the late-time light curve of SN~1994I, all of which is distributed
above $3,000\,\kms$ to avoid overestimating the flux around day $\sim\!40$. 
Both models need a substantial amount of $^{56}$Ni between $8000$ and
$10\,000\,\kms$ to reproduce the fast-rising early light curve 
($\sim\!0.008 \msun$ for CO21 and $\sim\!0.003 \msun$ for our model). This is
consistent with the blue nature of the early spectra, which indicates
substantial $\gamma$-ray heating and {\nifs} mixing in the outer part of the
ejecta (see \S\ref{sec:disc-vph}).

The light curve of our ejecta model reproduces the observed light curve of
SN~1994I well, both in the peak phase (before day $\sim\!30$) and in the tail
phase (after day $\sim\!50$).  By contrast, the light curve of CO21 declines
too fast to follow the observations after day $\sim\!50$ and the discrepancy
increases to $\sim\!50$ per cent in flux by day $\sim\!100$.  This confirms the
results of our nebular spectrum modelling that, compared to CO21, more
low-velocity mass is needed to generate the observed line luminosity in the day
$159$ spectrum. The low-velocity dense core may suggest that the explosion of
SN~1994I was actually aspherical. Spherical explosion models can not produce
this dense inner core because they exhibit a cutoff velocity below which
material falls back to form the compact object in the center of the explosion. 
In contrast, models of aspherical explosions generally show a higher density in
the inner, low-velocity core because the fall-back region is distorted
\citep[e.g.,][]{khokhlov99,maeda02,maeda03,mazzali04,tomita06}. Our model light
curve overestimates the flux by $<\!20$ per cent at the transition phase
between day $\sim\!30$ and day $\sim\!50$. Further tuning the $^{56}$Ni
distribution and the ejecta density structure in the low-velocity core could
solve this discrepancy, which is, however, not significant. On the other hand,
this may also be a multi-dimensional effect, the modelling of which is then beyond the
scope of this paper.

\section{Conclusions}\label{sec:conclusion}

We have re-investigated the properties of the Type~Ic supernova SN~1994I
because this object is among the best observed ``normal'' {\sneic} and possibly
represents a link between different branches of core-collapse supernovae. The
study was based on the density structure of the explosion model CO21 by
\citet{iwamoto94} although we allowed the composition to deviate from the
prediction of this model. Overall the explosion model seems to give a
reasonable representation in the outer parts of the ejecta, although the
mismatch of the velocities of some features in the early phase may suggest that
a density structure with a steeper gradient in the intermediate-velocity range
could be more suitable. The nebular-phase models, in contrast, require a
significantly larger mass at low velocities than predicted by the model. Our
estimate for the ejecta mass inside a radius corresponding to $v=5500\,\kms$ is
at least $0.4\msun$. From the photospheric model at $t=24\,$d, which has a
photospheric velocity of $5400\,\kms$, one can estimate the mass above the
outer velocity of the nebular spectrum to be about $0.6\msun$. This adds up to
a value that is larger than the total ejected mass of $0.9\msun$ of model CO21.

The spectral models also give some indication that the explosion of SN~1994I
exhibited some extent of asymmetry or at least inhomogeneity which was able
to mix {\nifs} out into the oxygen-rich layers. This is also supported by the
trend that the abundance of Fe-group elements actually seems to decrease toward
the centre of the ejecta. In turn, oxygen-rich material appears to be present
at very low velocities as can be seen in the nebular spectrum. In comparison
model CO21 does not show much mixing.  Recent hydrodynamical simulations
\citep{burrows96,lai00,scheck04} and spectropolarimetric observations
\citep{wang01,wang03a,leonard02,leonard06} of core-collapse supernova
explosions suggest the presence of global asymmetries which could induce large
mixing of the ejecta.
The overall picture for SN~1994I is that it probably originated from the
explosion of a low-mass core that stripped its envelope possibly by binary
interaction. The peculiar shape of the early-time spectra, as well as the very
fast decline of the light curve show that this object was by no means a
``typical'' {\snic}.  This should be kept in mind when comparisons to other
objects are made.

The amount of extinction of SN~1994I by dust is subject to large uncertainties. 
Our models suggest that the value of $\ebv=0.45\,$mag used by \citet{iwamoto94}
to model the light curves is probably significantly too large. A value of at
most $0.30\,$mag seems more suitable to fit the observed spectra.

Given the constraints on the ejecta mass and reddening from the spectral
models, we recalculated the theoretical light curve for the CO21 model with a
more massive low-velocity core region.  This new model still fits the observed
bolometric light curve that we constructed from the spectra well (including a
correction for the unobserved infrared bands). To obtain this fit it was
necessary to place a substantial amount of $^{56}$Ni at velocities between
$8000$ and $10\,000\,\kms$. This supports the corresponding conclusion from the
spectral fits. The total amount of $^{56}$Ni was $\sim\!0.07\msun$ and did not
change compared to the original CO21 model. (The effect of the lower reddening
than assumed by \citet{iwamoto94} is balanced by a larger distance.)

The nature of the IR-absorption around $10\,300\,${\AA} cannot be revealed
unambiguously on the basis of the methods chosen here.  The most likely
identification seems to be {\ion{C}{i}} at intermediate velocities with a
contribution of \ion{He}{i} at high velocities. Neither ion seems to be able to
account for the entire absorption without causing unseen features in other
parts of the spectrum. \ion{C}{i} can also account for other line features
which are not reproduced by our models otherwise. A more in depth analysis
using a consistent non-LTE treatment of all important processes including
non-thermal excitation of \ion{He}{i} is needed to provide a more conclusive
picture.

\section*{acknowledgements}

We acknowledge B.~Schmidt for providing us with some data and S.~Taubenberger
for re-calibrating the spectra. This work was supported in part by the European
Union's Human Potential Programme {\em ``Gamma-Ray Bursts: An Enigma and a
Tool,''} under contract HPRN-CT-2002-00294. D.N.S. thanks W.~Hillebrandt at the
Max Planck Institute for Astrophysics for his hospitality.  Some of the authors
thank the Kavli Institute for Theoretical Physics at the University of
California, Santa Barbara for its hospitality during the program on ``The
Supernova Gamma-Ray Burst Connection,'' where this paper was completed. This
research was supported in part by the National Science Foundation under Grants
PHY99-07949 and AST-0307894.
\bibliographystyle{mn2e}
\bibliography{jabref}

\end{document}